# The Open Autonomy Safety Case Framework

**Michael Wagner and Carmen Carlan**

Edge Case Research Inc., Pittsburgh, United States

## Abstract

*A system safety case is a compelling, comprehensible, and valid argument about the satisfaction of the safety goals of a given system operating in a given environment supported by convincing evidence. Since the publication of UL 4600 in 2020, safety cases have become a best practice for measuring, managing, and communicating the safety of autonomous vehicles (AVs). Although UL 4600 provides guidance on how to build the safety case for an AV, the complexity of AVs and their operating environments, the novelty of the used technology, the need for complying with various regulations and technical standards, and for addressing cybersecurity concerns and ethical considerations make the development of safety cases for AVs challenging. To this end, safety case frameworks have been proposed that bring strategies, argument templates, and other guidance together to support the development of a safety case. This paper introduces the Open Autonomy Safety Case Framework, developed over years of work with the autonomous vehicle industry, as a roadmap for how AVs can be deployed safely and responsibly.*

## 1 Introduction

The desire to use safety cases as safety communication concepts for commercial autonomous vehicles (AVs) is increasing, especially after the publication of UL 4600. A safety case is defined by UL 4600 as a "*structured argument, supported by a body of evidence, that provides a compelling, comprehensible and valid case that a system is safe for a given application in a given environment*". Arguments in safety cases are usually inductive, meaning that the truth of its premises provides some grounds for its conclusion. Safety case approaches support the flexibility needed by safety processes for autonomous vehicles to address novel technologies and utilise emerging safety strategies without prescriptive guidance from standards or accepted practices (Koopman et al. 2019).

AVs are systems consisting of components implementing complex algorithms, sensors, and machine learning (ML) components. The development and management of safety cases for AVs is challenging. A safety case for an AV shall argue about the confidence that all relevant mishaps, the causing hazards, and the causal factors of the hazards are identified and how the implementation of mitigations is sufficient to mitigate the risk of the identified hazard. To this end, a correct understanding of the behaviour and performance of such components and their interaction is needed. Whereas a safety case is an argument about the safety of a system in a given operating environment, AVs will operate in real-world environments, which are complex and unpredictable. This means that the validity and even soundness of the safety case may change over time. Arguments in safety cases are valid if the premises provide enough confidence that the conclusions are true. Valid arguments are sound when the premises are true. The development of AVs often uses novel technology, which is yet to be standardised. Developing compelling arguments supported by evidence generated using novel technology may impose specific challenges. Further, the safety case of AVs shall demonstrate compliance with a large set of standards and regulations. Lastly, a safety case of an AV shall also discuss safety-related cybersecurity and ethical considerations.

Safety case frameworks facilitate the creation of safety cases and guide their management. A safety case framework (SCF) captures the broad guidance necessary to develop a safe system and an adequate safety case. Apart from a templated argument structure, an SCF entails supporting concepts and strategies that guide the expansion of the argument, templates for supporting evidence, process definitions, and requirements checklists.

This paper introduces the Open Autonomy Safety Case Framework (OASCF), developed within Edge Case Research (ECR) over years of work with the autonomous vehicle industry, as a roadmap for how you can deploy safely and responsibly. Further, OASCF draws on Edge Case's experience writing and applying standards such as UL 4600, MIL-STD-882E, ISO 21448:2022, and ISO 26262:2018, as well as supporting developers of safe autonomous trucks and cars and assessing the safety of complex defence systems.

The OASCF is an overarching strategy and development philosophy for complex autonomous systems in the automotive industry. The goal of OASCF is to support safety engineers in developing a safety case in compliance with UL 4600 that argues that the system is developed in compliance with relevant standards mentioned by UL 4600, such as ISO 26262 and ISO 21448. Whereas the framework scopes automotive systems, UL 4600 also recommends following best practices from other industries, such as aerospace (Federal Aviation Administration (FAA) documents) and military (MIL-STD-882E). OASCF establishes how to implement a safety management system, communicate risks to stakeholders, and navigate relevant standards and regulations. The framework also enables effective, independent assessment processes defined in UL 4600. OASCF is *open*, meaning that it is implementation agnostic. OASCF can be used in different projects, implementing AVs for different use cases. Further, its openness is given by the fact that, hereby, we make the first nine levels of the templated high-level argumentation strategy public, because we believe that safety should be a prerequisite for deploying AVs, and not a competitive advantage. Therefore, we want to support the AV community to take the first steps in planning their safety programs.

The goals of OASCF are to:

- Enable deployment of safer autonomy by providing adequate safety management processes that developers can quickly adopt;
- Engender trust in safe autonomy by enabling AV developers to have a fully transparent safety case backed by clear evidence and a conformance monitoring plan;
- Drive effective regulation of autonomous vehicles; and
- Accelerate insurance and risk rating for fleets of autonomous vehicles.

## 2 Background on Safety Cases

The concept of safety cases was introduced thirty decades ago when, in 1989, the Control of the Industrial Major Accident Hazards in the British chemical industry mandated the generation of a written report named *safety case*, arguing about the mitigation of the hazards and risks of a site.

A *system safety case* is a well-constructed, easily understandable, and valid argument structured around ensuring the fulfilment of safety goals for a specific system operating within a defined environment. The safety goals are measures to mitigate the risk associated to the identified system hazards. The argument within the safety case is substantiated by compelling evidence. From a structural standpoint, a safety case comprises three fundamental components: safety claims regarding the system in question, a compilation of supporting evidence (such as safety analyses, software inspections, or functional tests) generated throughout the safety engineering lifecycle, and a rationale outlining how the available evidence contributes to demonstrating the achievement of the safety assertions. The claims can be further subdivided into sub-claims until each sub-claim is directly substantiated by evidence. The argument systematically captures the logical connections between claims, sub-claims, and evidence. The evidence supporting the argument is typically diverse, encompassing both quantitative and qualitative aspects and analytical and empirical data. Structured arguments may be graphically represented using different languages. One such language is the Goal Structuring Notation (GSN). Figure 1 shows an exemplar structured argument graphically represented using the Goal Structuring Notation (GSN). In GSN, the safety claims are represented via goals, the evidence via solution elements, and the rationale for how the safety claims are supported by evidence via strategies.

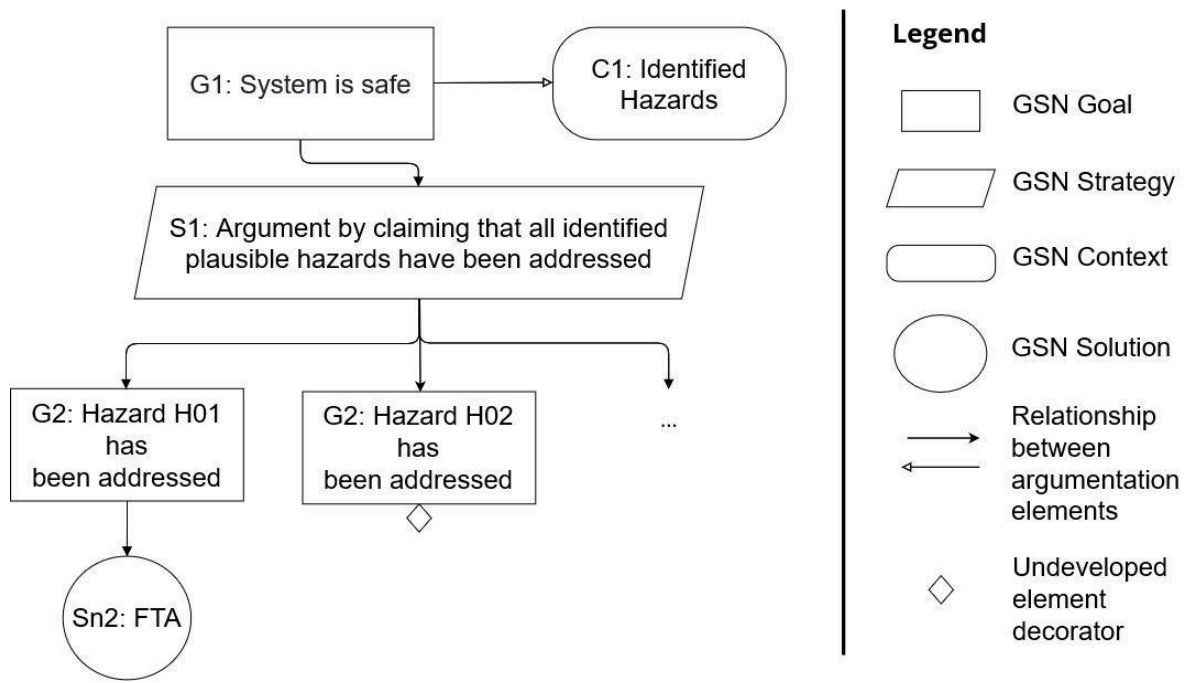

**Figure 1 ~ Exemplar Structured Argument in GSN**

Irrespective of the application domain, a safety case encompasses details about the system under evaluation and its operational environment. This includes defining the acceptable risks, identifying hazards, assessing the associated risks, specifying the adopted safety management system, and outlining responsibilities and the organisational safety policy.

In the automotive industry, Part 2 of ISO 26262 mandates that the system safety case compiles the system artefacts generated throughout the safety life cycle. Additionally, Part 10 of ISO 26262 mandates the presence of a plan for safety case maintenance in response to system changes. ISO 21448 offers examples of structured arguments concerning the achievement of Safety of the Intended Functionality (SOTIF). Whereas the Motor Industry Research Association's (MISRA's) Development Guidelines for Automotive Safety Arguments (MISRA 2019), a standard for software development in automotive systems, offers guidelines for articulating arguments related to software correctness.

UL 4600 outlines a set of best practices and considerations for developing system safety cases for autonomous systems. It supports safety engineers by prompting critical questions such as,

'Have you considered this?'. Moreover, UL 4600 emphasises the importance of analysing the impact of system changes on safety cases.

## 3  Live Safety Cases

It is not enough to build a single, static safety case that quickly becomes out of date once approved. Safety arguments that do not accurately reflect the current state of the system give a false sense of system safety. Safety cases should be considered living documents updated, reviewed, and published after each safety-relevant change to the system.

In compliance with UL 4600, the OASCF is based on continuous feedback loops to resolve sources of aleatoric and epistemic uncertainty in residual risk claims. Many of these feedback loops involve conventional incident reporting followed by hazard analysis and mitigation using level-of-rigour activities. Resolving uncertainty is especially important for applications of autonomous vehicles in open-ended use cases in the real world. This fact is compounded by novel technologies such as machine learning. The safety cases created with OASCF are intended to be live (see Figure 2).

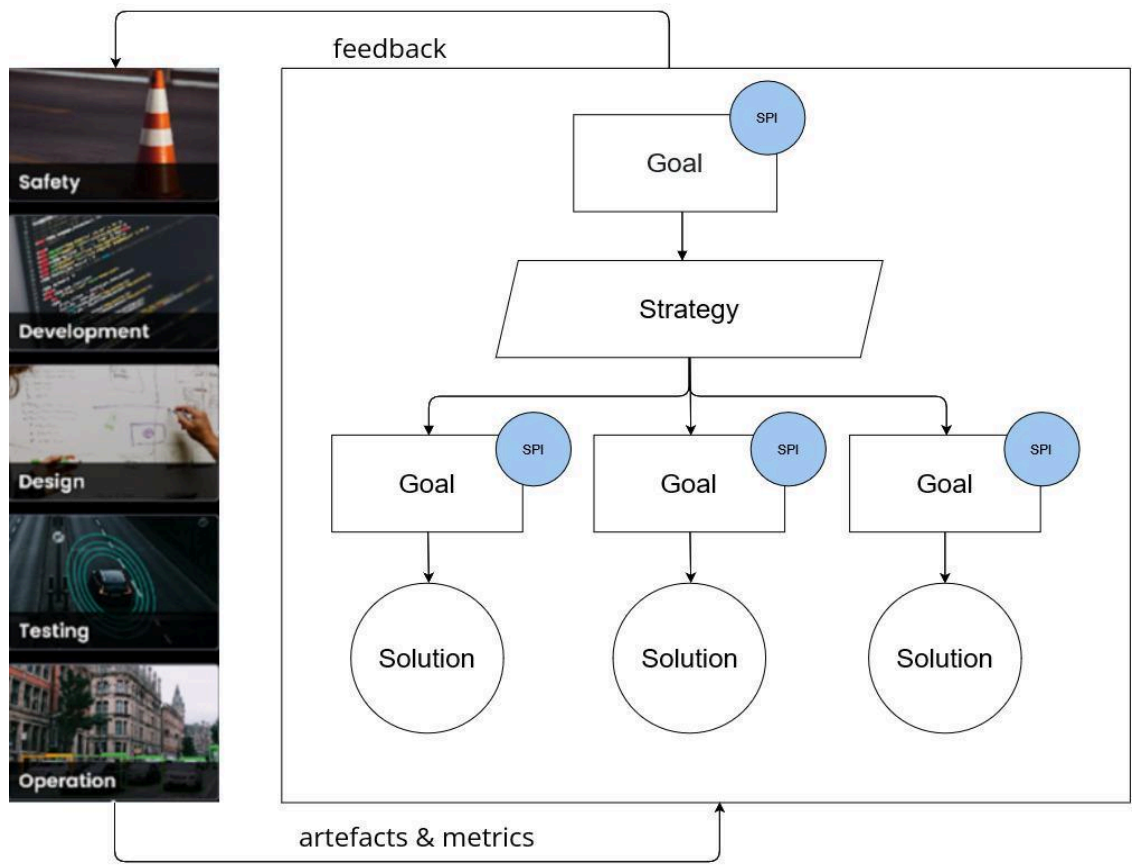

**Figure 2 ~ Overview of the concept of live safety cases, which have dynamic links to system artefacts, and which have SPIs attached to the claims to evaluate their soundness**

In comparison with regular safety cases, live safety cases have a set of additional properties:

- The truth of the claims in live safety cases may be automatically checked based on the existing evidence;
- Live safety cases are continuously fed with the most current safety evidence; and

- Live safety cases are sensitive to system changes. Whenever a system change occurs, the impact of the change on the safety case is automatically detected.

To this end, the claims in our OASCF are tied to templated Safety Performance Indicators (SPIs). While evidence and SPIs within the OASCF involve metrics data and thresholds, their purposes differ. Evidence supports the truth of a claim and is generally collected prior to an assessment. In contrast, SPIs detect situations where claims might be false. They are central to an assurance case monitoring plan defined by UL 4600. Templated SPIs can be tailored to data sources tied to a user's functional architecture and safety management system implementation.

Behavioural SPIs trace to autonomy functions such as perception, prediction, planning, and control. Operational SPIs trace to safety management systems, processes, and safety culture. SPIs are of two types: lagging and leading. Lagging SPIs directly measure the residual risk during system operation. Examples of lagging SPIs include fatalities, other crashes, violations of traffic rules, and near hits. Lagging SPIs are associated with claims higher up in the safety case and are measured during operation. Confidence in lagging SPIs increases as the deployment scales and, consequently, the volume of available data grows. Leading SPIs are measured both earlier in the system development lifecycle, including during simulation and road testing, and during deployment and operation, and may be used to predict system safety. Examples of leading SPIs include malfunctions, failures of system components, rates of sensor malfunctioning, perception failure rates, and software component execution faults. Leading SPIs are associated with claims deeper in the safety case. Leading SPIs may be used to identify so-called "triggering conditions" specified in ISO 21448, which identify operational situations in which an autonomy algorithm fails to mitigate hazards as intended. Figure 3 shows a fault tree that models how the occurrence of triggering conditions can lead to a loss event, such as a collision with a pedestrian. Confidence in the predictive nature of SPIs is to be gained by continuously comparing the predicted data with the actual data.

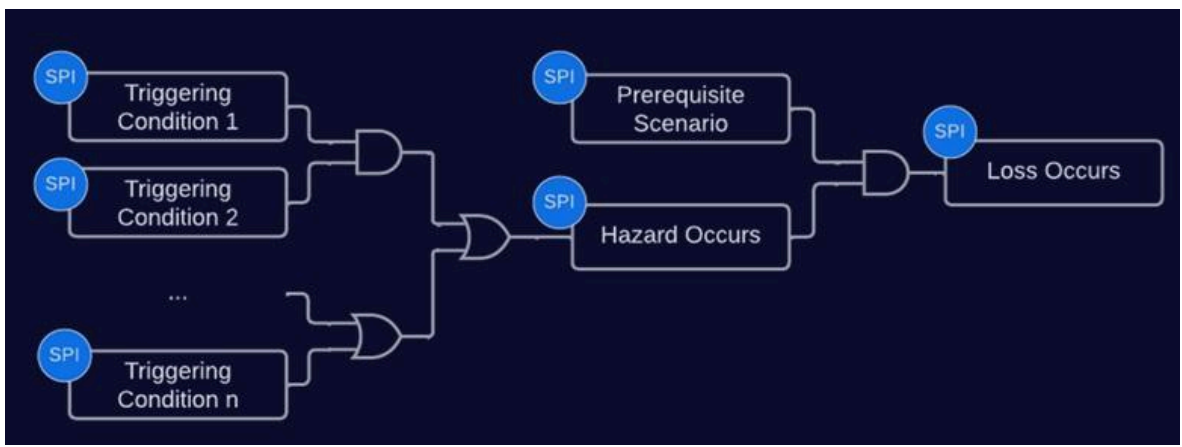

**Figure 3 ~ SOTIF Fault Tree Annotated with SPIs**

During system development, SPIs may detect the hazards whose desired risk reduction has not yet been achieved via the system design. One countermeasure could be to apply procedural mitigations to allow system maturation while maintaining safety. At the transition into operational deployment, all SPIs shall be satisfied. However, because of a lack of sufficient data, the confidence in the satisfaction of the SPIs is low. Post-deployment monitoring of SPIs aims to increase this confidence. Also, post-deployment, SPIs can detect violations of claims and assumptions in the safety case that invalidate the argument. Violations of the safety case detected in operation should trigger remediation processes, including potentially grounding the fleet while the causes are diagnosed, and the system or safety argument is updated.

Arguments in live safety cases shall be supported by evidence system artefacts, such as simulation test results, feedback from fleet vehicles, and dynamic hazard tracking mitigations referenced via dynamic links. Dynamic links allow the safety case to always reference the latest version of the system artefacts.

Given changes in the referenced evidence, the impact of those changes on the safety arguments needs to be analysed. Live safety cases shall also support automated safety argument change impact analysis (CIA) and shall be structured in a manner that reduces the impact of foreseeable changes. Given a change in a part of a safety case or in a referenced system artefact, the impact of the change on the elements of the safety case shall be automatically identified. Several approaches for automated safety argument CIA have been proposed in the literature. In Cârlan et al. (2022), we proposed an approach for a sound, automated safety argument CIA based on semantically rich annotations of dynamic links between argument elements and system artefacts, and of traces between argumentation elements.

Live safety cases have many benefits, including:

- Tracking and communicating development progress internally;
- Confidently communicating residual risk externally;
- Automating data flows with standard and custom data connectors;
- Creating and managing interdependencies between complex sets of claims;
- Tracking hazards, mitigations, and requirements sets;
- Monitoring for SPI violations;
- Providing continuous risk management updates; and
- Identifying edge cases that need to be addressed.

## 4 The Proposed Safety Argumentation Strategy

### 4.1 Preface

Our proposed OASCF provides its user with an overarching strategy and development philosophy toward a safe system and a compelling safety case including the following:

- A safety assurance process based on relevant safety standards;
- A set of techniques and templates for the generation of system development artefacts that are to be used as safety evidence and checklists for their assessment;
- A templated top-level argumentation strategy for why the system is safe, based on the proposed safety process, from which ANSI/UL 4600 compliant safety cases can be built;
- Assessment criteria for a well-formed and sound safety case based on the state of the art and best practices, such as ISO/IEC/IEEE 15026-2:2022, which specifies minimum requirements for the structure and contents of an assurance case;
- A catalogue of patterns for developing lower-level product-based, process-based, and confidence safety arguments;
- A process for continuously monitoring the status of the safety claims via Safety Performance Indicators (SPIs); and
- A catalogue of templated SPIs, mapped to the claims in the templated argumentation strategy.

In this paper, we only elaborate on the templated top-level argumentation strategy.

The OASCF provides an ecosystem of stakeholders with a roadmap for deploying safe autonomy. The OASCF may be used for safety planning, safety process execution, and safety risk decisions through development, deployment, and sustained system operation. An ecosystem adopts the framework by tailoring it to its risk acceptance criteria[1], deployment goals, autonomy technology, use cases for autonomous vehicles, and regulatory expectations.

The intended audiences for the OASCF are diverse, ranging from machine learning developers to insurance underwriters. These audiences depend on a consistent understanding of residual risk. Many organisations unfortunately juggle disparate views about risk even across their own internal teams. Worse still, regulators responsible for upholding public safety often receive incomplete and optimistic views of risk. The OASCF seeks to bridge these gaps by providing a coherent assessment of residual risk viewed through lenses suitable for different audiences. The template for the structured argument shows how system artefacts generated by different teams support the risk assessment together. Further, the OASCF supports the definition and evaluation of safety-relevant tasks and artefacts throughout the development process, as it includes templates and evaluation criteria for evidence artefacts, as well as the engineering processes used to produce the artefacts.

OASCF harmoniously combines best practices from the following sources into a single compelling argument that is supported by evidence generated by a unified safety engineering process:

- Safety case assessment from ANSI/UL 4600;
- Safety management systems from the FAA (2020);
- Hazard analysis from MIL-STD 882E;
- Engineering rigour from standards relevant for users across industries, such as the Joint Software Systems Safety Engineering Handbook (USDoD 2010)., or ISO 26262;
- Validation of the suitability of autonomy algorithms from ISO 21448 (SOTIF);
- Monitoring of safety performance indicators from UL 4600; and
- Lifecycle safety processes best practices.

## 4.2 Positive Trust Balance

Autonomous vehicles that rely on machine learning for life-critical functions like tracking pedestrians and surrounding traffic hold the promise of being safer than human drivers. To this end, Positive Risk Balance (PRB) shall be demonstrated. Demonstrating the achievement of PRB before the deployment of autonomous vehicles without sufficient lagging metric data to provide high confidence in an acceptable safety outcome is difficult. The argument employed by the OASCF is inspired by (Koopman and Wagner 2020) and supports a responsible deployment decision despite the lack of lagging metric data. This approach relies on practical evidence collection, supporting the expectation of sufficiently low risk rather than the unrealistic requirement of conclusive proof that a risk target has been met right from the outset. Positive Trust Balance (PTB) proposes using a combination of validation, engineering rigour, post-deployment feedback, and safety culture to predict deployment risk confidently. Further, responsible deployment requires monitoring key assumptions and safety metrics post deployment and to continuously calibrate risk.

The top-level claim of our templated high-level structured argumentation strategy is "*The autonomous driver is safe enough to operate in the considered operational design domain*". An autonomous driver is a system embedded in a vehicle, which navigates and operates the vehicle without human intervention. An autonomous driver uses various sensors, such as cameras, radars, and lidars, to perceive the operating environment of a vehicle, make decisions, and perform driving tasks without human input. Autonomous drivers can have different levels of automation, ranging from Level 0 (no automation) to Level 5 (full automation). Implementing the PTB approach, the OASCF relies on three pillars of argumentation (see Figure 4). The first pillar argues about having confidence in the safety argument given the employed processes in the organisation. The second pillar argues that the system is safe by design, namely that the system was engineered and tested using rigorous processes while considering the identified hazards. The third pillar argues that the hazards uncovered during system operation are appropriately addressed.

---

[1] Risk acceptance criteria is defined in ISO 21448 as a criterion representing the absence of an unreasonable level of risk. Acceptance criteria may be quantitative or qualitative. One example of such criterion given in the standard is the maximum number of incidents per hour. What is considered 'unreasonable' depends on the organisation's risk appetite.

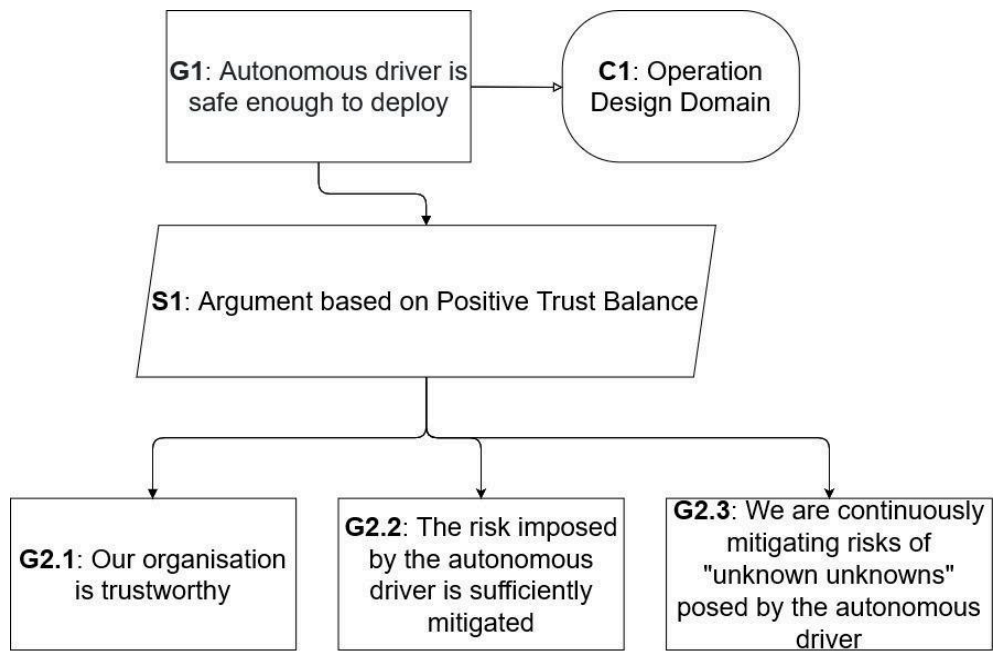

**Figure 4 ~ Top-level Claims of OASCF in GSN**

### 4.3 Live It Right

OASCF requires a robust and transparent safety culture to ensure sound technical outcomes and build public trust over time. The "Live It Right" pillar (see Figure 5) argues that the organisations building and operating the AV are safe. This means that these organisations have strong safety management and safety culture and that they deploy novel technology responsibly. Establishing this allows the reader of the safety case to trust the rest of the argument.

This section of our framework covers critical issues such as:

- Avoiding setting unreasonably low initial quality and validation goals based on an argument that post-deployment updates will fix bugs. Such an argument is not aligned with the Positive Trust Balance approach. Instead, risk acceptance criteria must be aligned with industry best practices, regulatory requirements, and societal expectations. Our framework uses guidance from the German Ethics Commission (BMVI 2017) on setting nuanced quantitative risk acceptance criteria and safety integrity level approaches from safety standards.

- Using the lack of maturity in accepted practices in some areas (e.g. still evolving best practices for safe machine learning) as an excuse for not following well-known best practices for more traditional aspects of the system, such as functional safety.

Risk acceptance criteria, organisational principles, and safety management plans form the basis of a safety management system (SMS) safety policy and safety promotion processes. SPIs tied to "Live It Right" claims track deviations from these policies and processes.

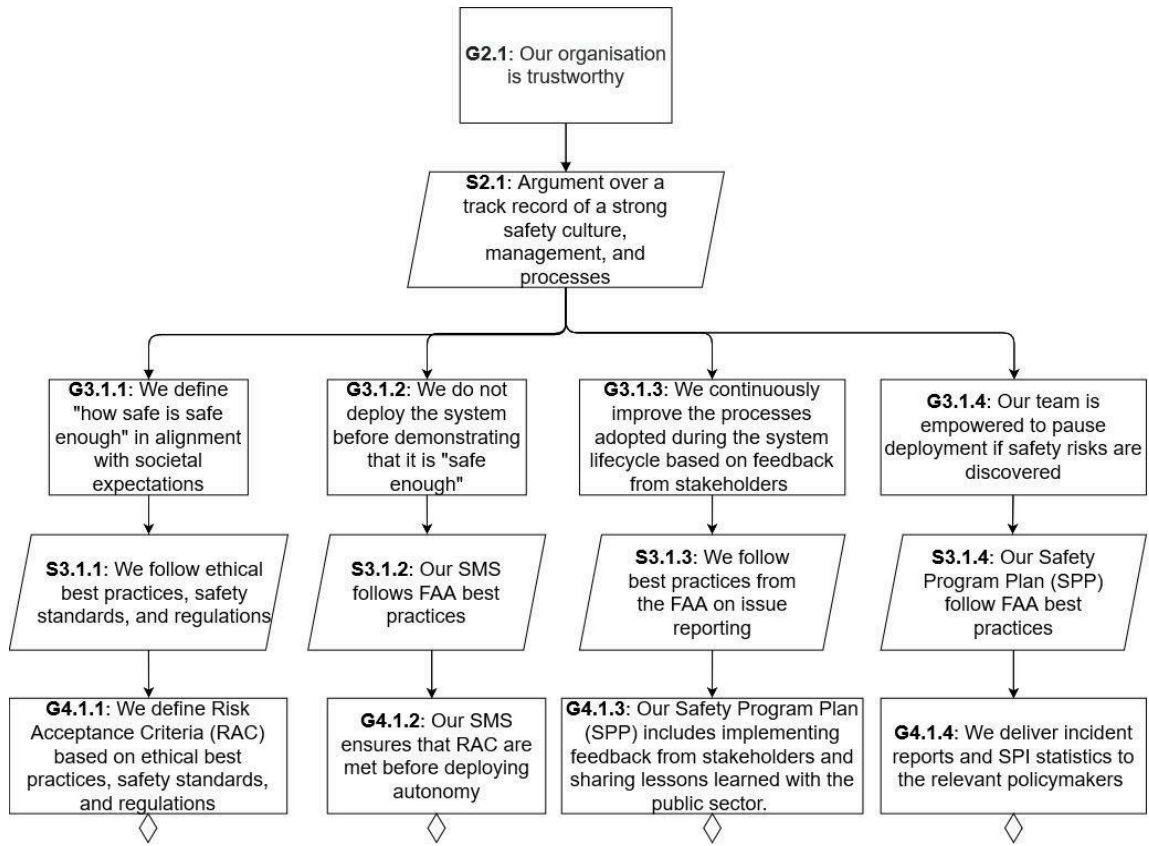

**Figure 5 ~ The "Live It Right" Argumentation Pillar**

## 4.4   Engineer It Right

This part of the safety case (see Figure 6) argues that the autonomy technology is safe by design, and the development and safety assurance of the system is grounded on rigorous engineering development practices. The OASCF demands that the manufacturer of the autonomous system builds their technology in accordance with industry standards for identifying and mitigating safety risks. Defence standards, like MIL-STD-882E, refer to this as a system safety process, whereas the FAA refers to it as safety risk management. These standards define how to:

1. Identify loss events that can occur within a defined concept of operations;
2. Assess and track hazards that could lead to these loss events and understand how the autonomous vehicle can cause these hazards can occur;
3. Define safety-critical functions that prevent hazards from occurring;
4. Implement safety-critical functions with a level of rigour appropriate to the severity of risk being mitigated;
5. Verify that implementations are correct, following the guidance from standards; and
6. Validate that hazards have been mitigated as intended in operation, following the guidance from standards.

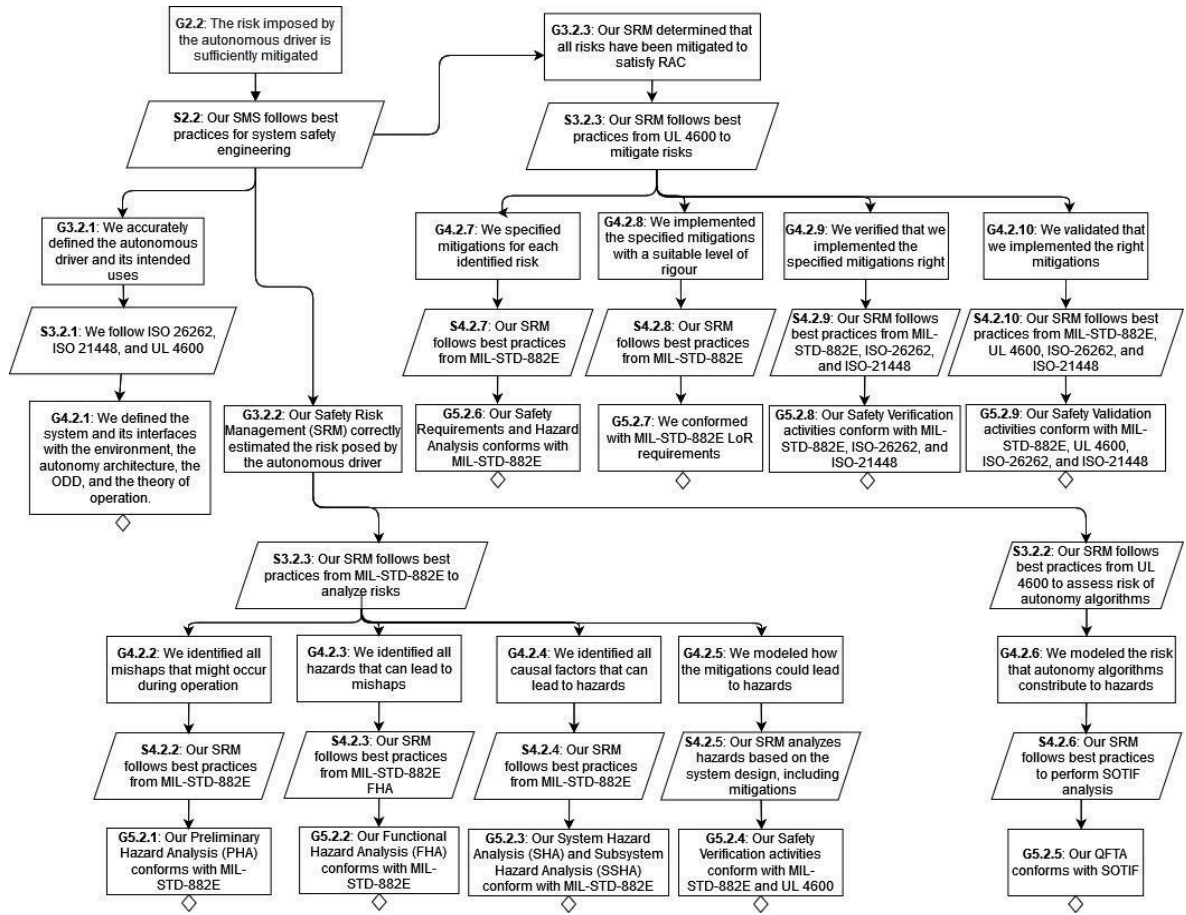

**Figure 6 ~ The "Engineer It Right" Argumentation Pillar**

Safety standards provide tables specifying methods for the above tasks and the work products they generate. The OASCF relies on all these work products and independent assessments of the manufacturer's conformance with relevant standards. Our framework provides templates for each of these work products. However, while we publish the OASCF, we do not intend to publish the templates.

Among others, quantitative fault tree analysis (QFTA) generates the evidence supporting this argumentation pillar. For each identified hazard, the safety case refers to a quantitative fault tree (QFT) that models our best understanding of how causal factors can lead to a hazard, along with expectations for how often these causal factors will occur in practice. The QFT consists of a function (typically Boolean) describing how the combination of conditions referred to as causal factors can lead to a hazard.

The structure of a QFT is generated by well-known hazard analysis techniques such as those used in MIL-STD-882E. First, a preliminary hazard analysis (PHA) defines what loss events can occur and lists what hazards can lead to these loss events. Second, the functional hazard analysis (FHA) allocates the occurrence of hazards to functions in the autonomy stack and builds the fault tree. Third, the system/subsystem hazard analysis process allocates functions to components in the architecture, and can employ techniques such as failure modes, effects, and criticality analysis (FMECA) to define further what causal factors need to be tracked, including hardware failures and software defects.

Risk targets or estimates can be allocated to each node in the fault tree. Initially, this can be used to define a risk budget for the autonomy stack and operations. However, risk can be estimated

— and the risk budget refined — with feedback from SPI monitoring during simulation and on-road testing. During initial deployment, statistical deviations between SPIs seen in validation and those seen in real-world operations can serve as a trigger for an issue for the safety management system to resolve. Further, we use the QFTs to define safety requirements on the operations of an autonomous vehicle in a way that can be monitored using telemetry, examined for anomalies and gaps in hazard analysis, and used to improve the safety case over time.

Mitigating hazards will also require instituting operational controls. Reasons for this include the need for machine learning in safety-critical functions, the open-ended nature of requirements on autonomous behaviours, and unavoidable uncertainty in our models of the environments in which autonomy operates over time. An understanding of all such situations is the aim of methods in MIL-STD-882E and ISO 21448. For example, in automotive, operational control might include limitations on driving routes that stay away from school zones, bike lanes, and densely populated urban areas. Other operational controls require preventative maintenance and diagnostics to ensure that safety-critical functions in the autonomous vehicle are dependable. While testing autonomous vehicles, operational controls usually include safety drivers with dependable takeover mechanisms outside the scope of the autonomous vehicle itself, for example, as defined in SAE J3016.

## 4.5   Operate It Right

"Operate It Right" is the argument that the operation of the autonomous driver is safe (see Figure 7). The claim that an autonomous system is "operated right" is based on two assertions: that all the operational controls are correctly in place, and that the QFTA accurately describes the rates of causal factors that occur in practice.

Together, these two claims build an argument that the safety assurance processes of an SMS is effective. An SMS uses safety assurance processes to evaluate the continued effectiveness of mitigations in the field, and to identify previously unknown risks and hazards.

Process audits can provide evidence that operational controls are properly in place. However, audits measure inputs to the operating system, not the system's outputs. Consequently, OASCF also demands evidence from safety performance indicators, or SPIs, that monitor the behaviour of the autonomous driver in operation.

The data from the operation are recorded, analysed via SPIs, and acted upon by the manufacturer to assure safety in the field. SPIs are central to the FAA's safety assurance strategy, and are also used by UL 4600 to monitor safety case assumptions in real-world operation. The analysis of SPIs supports the following goals:

- Identifying areas where safety performance can be improved;
- Evaluating the effectiveness of operational safety risk controls;
- Increasing confidence in the system to validate the expected risk of operation; and
- Minimising harm by identifying unknown issues and risks before losses occur.

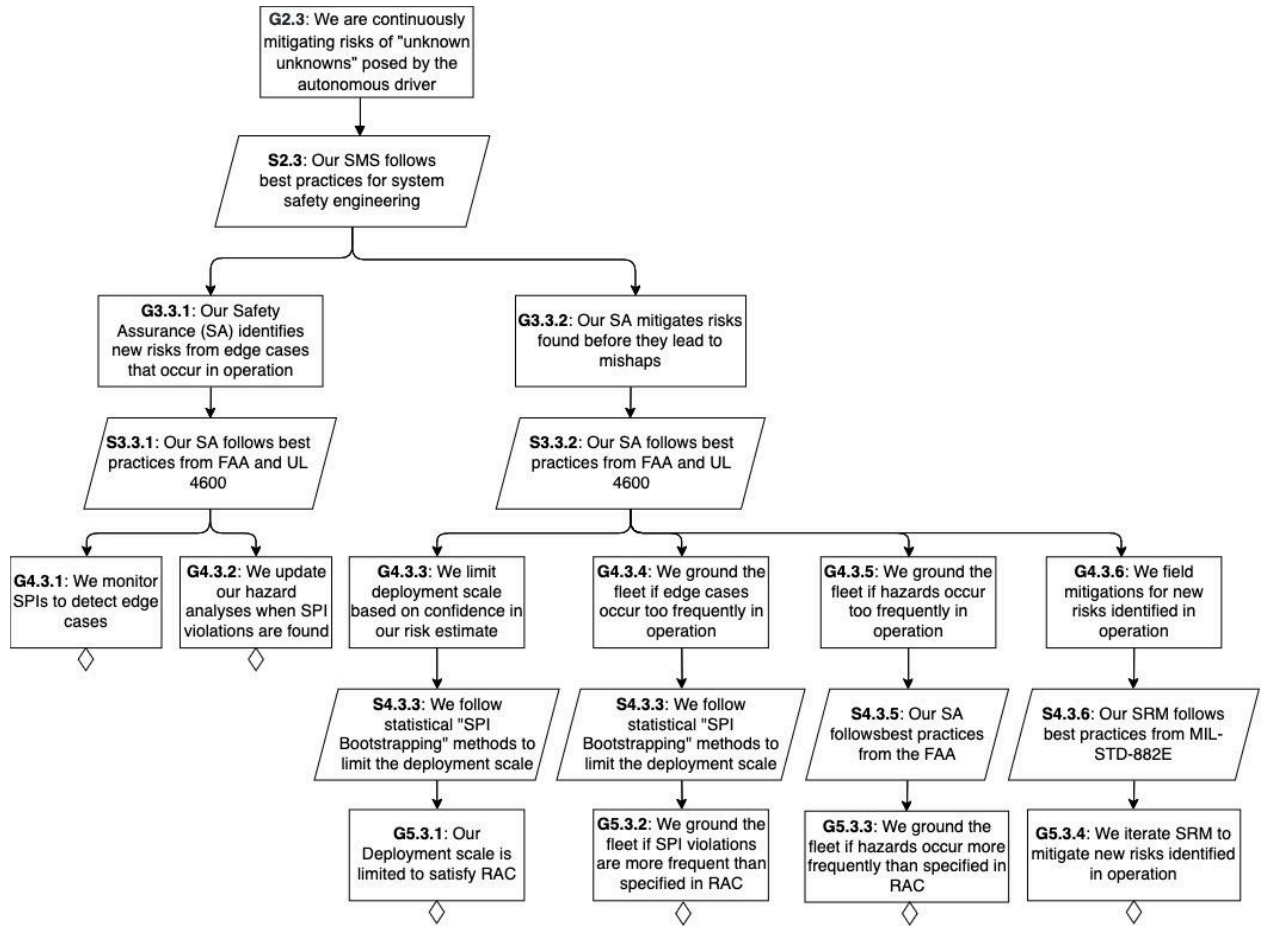

**Figure 7 ~ The "Operate It Right" Argumentation Pillar**

# 5 Related Work

## 5.1 Safety Cases for Autonomous Systems

Several works in the literature have discussed how to structure safety arguments for AVs.

Guarro et al. (2017) propose a risk-informed safety case framework for unmanned aircraft systems, accompanied by a validation and verification framework. The safety case framework connects the risk-based claims with the evidence generated by applying the validation and verification (V&V) framework. This work focuses on integrating V&V results into the engineering-based argument but does not address broader safety argument concerns such as safety culture and post-deployment monitoring.

Hawkins et al. (2021) propose a process for the Assurance of Machine Learning in Autonomous Systems (AMLAS), accompanied by a set of GSN-based safety argument patterns. However, their proposed patterns argue that an ML-based component satisfies its allocated safety requirements. These safety argument patterns may be used to refine the claims in OASCF.

Wishart et al. (2023) propose a safety case framework for AVs, which argues that an AV is safe for its intended implementation based on three argumentation pillars. Like OASCF, their framework considers UL 4600, ISO 26262, ISO 21448, and different AVSC reports. The first two pillars — "SMS" and "Design Methods" — reassemble our "Engineer It Right" pillar, as they argue that an SMS and a safety engineering process based on standard requirements and best practices are in place during system development. The third pillar of the SCF proposed by Wishart et al. (2023) is "Scenario-Based Testing", arguing about a scenario-based testing process employed to evaluate the performance of the AV. This third pillar can be used to develop OASCF's Goal G5.2.8: "*Our Safety Verification activities conform with MIL-STD-882E*" further.

## 5.2    Safety Case Frameworks for Autonomous Vehicles

As shown in the following, our OASCF aligns with existing SCFs released by AV companies. Our OASCF can be seen as a *meta* SCF, which can be adapted by AV companies based on their specific needs, namely specific audience, use cases, used standards, and norms. Also, an adapted SCF shall be in line with the company's safety process. Once the OASCF is enhanced with company-specific information, it can be used to create safety cases for different systems.

The first organisation to use a safety case for self-driving cars was Uber ATG. Uber ATG published its safety case framework after an independent review by a team at ECR. This framework has since been adopted by Aurora Innovation, who acquired Uber ATG in 2021. The initiative taken by Uber ATG and Aurora has hastened the adoption of safety cases throughout the self-driving industry, prompting safety case development at Locomation, TuSimple, Waymo, Zenzic, and other organisations. Aurora (2023) published the first four levels of their high-level structured safety argument in a GSN format. The argument has five argumentation pillars, which can be mapped to our three pillars.

Aurora's, "*The self-driving vehicle is proficient during nominal operation*", "*The self-driving vehicle is fail-safe*" and "*The self-driving vehicle is resilient to reasonably foreseeable misuse and unavoidable events*" high-level claims map to the "Build It Right" pillar in OASCF, whereas their "*The self-driving vehicle and safety processes are continuously evaluated and improved*" maps to our proposed "Operate It Right" argumentation pillar. Our "Live It Right" argumentation pillar is reflected in Aurora's "*The enterprise is trustworthy*" claim.

The safety case framework proposed by Waymo has the absence of unreasonable risk as its top-level goal (Favaro et al. 2023). Waymo considers three types of hazards, namely architectural, behavioural, and operational. To assess the residual risk associated with the identified risks, risk acceptance criteria are defined. Similar to what OASCF proposes, the safety approach proposed by Waymo is dynamic and relies on safety as an emergent development property, safety as an acceptable prediction and observation, and safety as continuous confidence growth. This means that, similarly to OASCF, Waymo's framework requires that rigorous engineering practices are used during system development, actual and simulated performance is combined to measure safety performance indicators, and feedback loops exist to address safety issues occurring after deployment and to confirm the residual risk computed before deployment. The framework proposed a three-layered argumentation:

- Performant, secure, and robust hardware platform (drive-by-wire);
- Safe and responsible driving behaviour (autonomy); and
- Safe deployment and operations (operations).

OASCF's "Engineer It Right" argumentation pillar may be used to develop the first two layers, whereas the OASCF's "Operate It Right" pillar is one way to structure Waymo's third argumentation layer. To evaluate the evidence and the arguments within the safety cases generated while using Waymo's Safety Case Framework, Waymo proposed a Case Credibility Assessment. OASCF instead proposes templates and checklists, for generating and evaluating credible arguments and evidence.

The Zenzic safety case framework (Zenzic 2021) supports the development of safety cases for AV testing, explicitly targeting the CAM Testbed in the United Kingdom. The framework provides its users with guidance on how to follow standards required for being able to test in the UK, such as ISO 26262, ISO 21448, PAS 1881:2020, PAS 1883:2020, while explicitly avoiding defining specific methods or tasks, instead providing useful guidance around the organisation and content of the safety case. While the Zenzic framework does not provide any concrete structured argument specified in a graphical notation, it proposes three argumentation strategies. One strategy argues about the functional safety and the safety of the intended functionality of the system under test. OASCF's "Engineer It Right" argumentation pillar may be used to construct a structured argument for this pillar.

The second argumentation strategy proposed by Zenzic framework concerns operational safety. It aims to demonstrate that a test vehicle can operate safely within the defined environment, relying on evidence that considers the interaction of the test vehicle with the operating environment, including the route, safety driver or operator, passengers and other road users. Here, OASCF's "Operate It Right" argumentation pillar may be used to construct a convincing, structured argument.

Zenzic's third argumentation strategy considers security threats, such as physical access or via electronic and telecommunications means (cybersecurity). Whereas OASCF does not yet consider security assurance, we plan to extend it with an argumentation strategy concerning cybersecurity in the future, as mandated by ISO/SAE 21434:2021. The Zenzic framework also advocates for safety cases to be live documents, and to be updated when previously-unknown hazards are uncovered. However, it does not propose a solution for how to specify a live safety case. Whereas OASCF is based on PTB, the Zenzic framework argues that the identified residual risk is As Low As Reasonably Practicable (ALARP)[2].

## 6 Discussion

**The need for specifying safety case interfaces**. The autonomous driver is usually embedded in the vehicle, usually receiving information from the sensors or sending commands to the actuators in the vehicle. This means that the safety case created for the autonomous driver, based on our proposed OASCF, shall be embedded in the safety case of the vehicle. Consequently, to ease the integration between safety cases, safety case interfaces shall be defined, specifying the assumptions and guarantees of a safety case related to, for example, the followed regulations and standards, the considered failure models, the achieved SIL/acceptance criteria and validation targets, the failure modes, or the safety requirements/safety goals/performance requirements.

**The challenge of implementing "live" safety cases**. The level of rigour needed for implementing live safety cases will take time to be achieved by autonomous driving development companies. Specifying SPIs and implementing an automated CIA are time-consuming and challenging tasks. This is why ECR has a catalogue of SPIs that can be customized for different projects, a toolchain, and a process supporting automated CIA. Further, to close/approve a live safety case before deployment, relevant stakeholders would also need to approve the defined SPIs and the implemented automated CIA. Another challenge is to specify a process for responding in case SPIs are invalidated while pondering between system availability and system safety.

---

[2] A term from UK safety legislation.

# 7 Summary and Future Work

In this paper, we introduced OASCF — a safety case framework for autonomous systems, based on the PTB rationale. OASCF has been developed based on ECR's vast experience gained while working with different stakeholders in the AV industry, from developers to regulators. Safety concepts and argument templates for autonomous vehicles quickly rely on implementation and organisation-specific details. A significant difficulty in generating generic argument templates is finding a balance between providing argument content and over-prescribing the details through templating. The templated structured argumentation within OASCF is high-level, focusing on processes, allowing program-specific details to be incorporated through references and evidence artefacts. Whereas the first nine levels from the OASCF we published in this paper showcase how to use various standards in order to build up a strong safety argument, the argument patterns accompanying the templated argument, which support safety engineers in further-developing templated arguments, showcase how to reference all work products mandated by the standards. Next, we plan to publish some of the argument patterns. In this paper, we discussed how our proposed OASCF covers the claims of existing SCFs, such as the ones proposed by Aurora, Uber, and Waymo. Our OASCF can be seen as a *meta* SCF, which different companies can instantiate.

Soon, we will apply OASCF for use cases and system types. Also, we plan on developing the SCF further so that it addresses European and American regulatory requirements. For example, European Regulations (EU) 2019/20144 and (EU) 2022/1426 pose explicit performance requirements for AVs/Automated Driving Systems (ADS). (EU) 2022/1426 also guides the derivation of scenarios relevant to the Operational Design Domain (ODD) of the ADS. In the United States, AV companies must satisfy licensing requirements, comply with specific safety standards, and follow testing protocols. Whereas our SCF guides its users towards compliance with a set of safety standards, we plan to enhance it to facilitate compliance with cybersecurity standards, such as SAE J3061_201601 and ISO/SAE 21434. Further, we will publish a catalogue of state-of-the-art safety argument patterns supporting the creation of arguments supporting our high-level templated argument. Also, another next step is to propose a catalogue of SPIs that, together with our OASCF can be used to specify live safety cases.

Our proposed SCF is *open*, meaning that everyone is invited to:

1. use our SCF to create their safety cases, and
2. contribute to perfecting the SCF so that all AV stakeholders can benefit from the safe deployment of AVs.

Hereby, we invite everyone to find assurance defeaters attacking our safety claims, argumentation, strategies, and referenced evidence and send them to us to enhance the OASCF.

## Correspondence Address

Michael Wagner: mwagner@ecr.ai

Carmen Carlan: ccarlan@ecr.ai